\documentclass{article}
\usepackage{spconf,amsmath,graphicx}
\usepackage{graphicx} %Loading the package
\usepackage{color}
\usepackage{tabularx}
\usepackage{booktabs}
\graphicspath{{Figures/}} 
\usepackage{multirow}
\usepackage{enumitem}
\setlist{nosep, leftmargin=14pt}

\usepackage{mwe} % to get dummy images
\definecolor{darkgreen}{rgb}{0,0.6,0.2}

\DeclareMathAlphabet{\mymathbb}{U}{BOONDOX-ds}{m}{n}
\makeatletter
\newcommand*\titleheader[1]{\gdef\@titleheader{#1}}
\AtBeginDocument{%
	\let\st@red@title\@title
	\def\@title{%
		\bgroup\normalfont\large\centering\@titleheader\par\egroup
		\vskip1.5em\st@red@title}
}
\makeatother
\title{
Demonstrating the risk of imbalanced datasets in Chest x-ray image-based diagnostics by prototypical relevance propagation}
\titleheader{Accepted at the 2022 IEEE 19th International Symposium on Biomedical Imaging (ISBI)}
\name{Srishti~Gautam$^{\star}$,  Marina~M.-C. Höhne$^{ \dagger\star}$,  Stine~Hansen$^{\star}$, Robert~Jenssen$^{\star}$ and
	Michael~Kampffmeyer$^{\star}$}
 \address{$^{\star}$ UiT The Arctic University of Norway, Tromsø, Norway\\
     $^{\dagger}$Technical University of Berlin, Berlin, Germany}

\begin{document}

\maketitle

\begin{abstract}
The recent trend of integrating multi-source Chest X-Ray datasets to improve automated diagnostics raises concerns that models learn to exploit source-specific correlations to improve performance by recognizing the source domain of an image rather than the medical pathology.
We hypothesize that this effect is enforced by and leverages label-imbalance across the source domains, i.e, prevalence of a disease corresponding to a source.
Therefore, in this work, we perform a thorough study of the effect of label-imbalance in multi-source training for the task of pneumonia detection on the widely used ChestX-ray14 and CheXpert datasets. The results highlight and stress the importance of using more faithful and transparent self-explaining models for automated diagnosis, thus enabling the inherent detection of spurious learning. They further illustrate that this undesirable effect of learning spurious correlations can be reduced considerably when ensuring label-balanced source domain datasets. 

\end{abstract}
\section{Introduction}
\label{sec:intro}

Current approaches for computer-aided diagnosis using Chest X-Ray images and deep learning tackle the lack of labeled data by leveraging data from multiple sources to achieve state-of-the-art performance \cite{miccai_combine}. However, the validity of this approach has recently been questioned by illustrating that models trained on datasets where each source exclusively contains labeled samples from a single class can directly learn the source peculiarities to solve the task \cite{DeGrave2021}.

\begin{figure}[!t]
    \centering
    \includegraphics[scale=0.42]{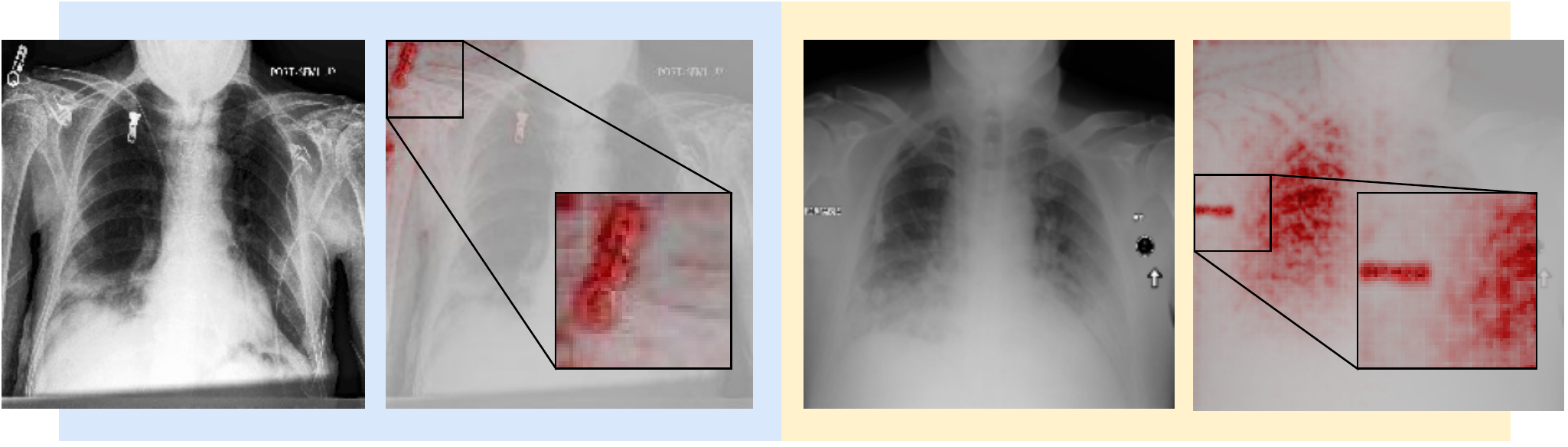}
    \caption{{
    Heatmaps of models for 90\% (blue) and 60\% (yellow) label-imbalance demonstrating spurious learning. With more imbalance, the reliance on the source annotations increases.}}
    \label{fig:tag_prototype}
\end{figure}

In this work, we illustrate that this behavior goes far beyond this extreme setting of source-based label exclusiveness where the models are prone to relying on spurious correlations even in the presence of minor imbalances of disease prevalence across the sources.
Specifically, the models tend to pick up on the textual image annotations present in the Chest X-Ray images, which may include metadata such as orientation or timestamp as well as information about patients, wards, and hospitals \cite{mimic_cxr}. This can lead to falsification of performance statistics as the model appears to be working well when in reality failing to capture class-related pathology-based characteristics. Our hypothesis is validated by performing a thorough analysis on the combination of the two commonly used Chest X-Ray datasets, ChestX-Ray14 \cite{chestxray14} and CheXpert \cite{chexpert} for the scenario of pneumonia detection, thus simulating two different sources of X-Ray images. We deliberately introduce a gradual imbalance in the prevalence of pneumonia images from one hospital system to assess the behavior of the model. Experimental results demonstrate that the model learns source related text-annotations (see Fig. \ref{fig:tag_prototype}).

This unanticipated behavior can go unnoticed with black-box models, thus advocating the use of explainable AI.
We, therefore, 
illustrate how these spurious correlations can be detected with the help of explainable methods. In particular, we rely on a self-explaining approach that can inherently explain the model's underlying decision strategies without relying on post-hoc approaches, thus generating more faithful 
explanations \cite{rudin2019stop}.
We leverage Prototypical Relevance Propagation (PRP) \cite{gautam2021looks}, a model-aware extension of the self-explaining model ProtoPNet \cite{protopnet} that provides more spatially accurate and high-resolution prototypical explanations, to further support our hypothesis.

With this work, we contribute to sharpen the
awareness for the use of label-balanced multi-source datasets as well as the importance of the use of self-explanatory models for computer-aided diagnostics.

\renewcommand{\tabcolsep}{2pt}

\begin{table}[!t]
\caption{Controlled test configurations for assessing behavior of recognizing source annotations. P and NP refer to Pneumonia and Non-Pneumonia class, respectively. The percentages denote the amount of images selected from H1 and H2. }
\def\arraystretch{0.9}
\begin{tabular}{|c|ccc|l|}
\hline
\multirow{1}{*}{\textbf{\small{Name}}}   & \multicolumn{3}{c|}{\textbf{\small{Setup}}}                                          & \multicolumn{1}{c|}{\multirow{1}{*}{\textbf{\small{Hypothesis}}}}                                                                                                                                                                                                                                            
% \\ \cline{2-4}
                                    %  & \multicolumn{1}{c|}{}   & \multicolumn{1}{c|}{\small{H1}}    & \small{H2}                    & \multicolumn{1}{c|}{}                                                                                                                                                                                                                                                                                
                                    \\ 
                                     \hline\hline
\multirow{3}{*}{\textbf{{\rotatebox[origin=c]{90}{\parbox[c]{1cm}{\centering \small{Test- 100H1}}}}}} & \multicolumn{1}{c|}{}  & \multicolumn{1}{c|}{\small{\textbf{H1}} }& \small{\textbf{H2}}                  & \multirow{3}{*}{\begin{tabular}[c]{@{}l@{}}\small{If the model relies only on the source}\\ \small{annotations, the accuracy of this is} \\   \small{expected to be 100\% for 100H1-0H2} \\ \small{label-imbalance and 0\% for 0H1-100H2.} \end{tabular}} \\
\cline{2-4}
& \multicolumn{1}{c|}{\small{\textbf{P}} }& \multicolumn{1}{c|}{\small{100\%}}   & \small{\small{0\%}   }              &
\\ 
\cline{2-4}
                                     & \multicolumn{1}{c|}{\small{\textbf{NP}} }& \multicolumn{1}{c|}{\small{0\%}}   & \small{\small{100\%}   }              &                                                                                                                                                                                                                                                                                                       \\ 
                                     \cline{2-4}
                                     & \multicolumn{1}{l}{}    & \multicolumn{1}{l}{}       & \multicolumn{1}{l|}{} &                                                                                                                                                                                                                                                                                                                  \\
                                     %\hline
                                     \midrule
\multirow{3}{*}{\textbf{{\rotatebox[origin=c]{90}{\parbox[c]{1cm}{\centering \small{Test- 100H2}}}}}} & \multicolumn{1}{c|}{ } & \multicolumn{1}{c|}{\small{\textbf{H1}} }  & \small{\textbf{H2}}               & \multirow{3}{*}{\begin{tabular}[c]{@{}l@{}}\small{The behavior of this setting is expected} \\ \small{to be the opposite to that of Test-100H1.}\end{tabular}} \\
\cline{2-4}
& \multicolumn{1}{c|}{\small{\textbf{P}} }& \multicolumn{1}{c|}{\small{0\%}}   & \small{\small{100\%}   }              &                                                                                                                                                       \\ \cline{2-4}
                                     & \multicolumn{1}{c|}{\small{\textbf{NP}}} & \multicolumn{1}{c|}{\small{100\%}} & \small{0\% }                  &                                                                                                                                                                                                                                                                                                                   \\ %hline
                                     \midrule
\multirow{3}{*}{\textbf{{\rotatebox[origin=c]{90}{\parbox[c]{1cm}{\centering \small{Test- 50-50}}}}}} & \multicolumn{1}{c|}{\small{}}  & \multicolumn{1}{c|}{\small{\textbf{H1}}}  & \small{\textbf{H2}}                 & \multirow{3}{*}{\begin{tabular}[c]{@{}l@{}}\small{The accuracy of this test set for different} \\ \small{ imbalances will indicate the learning}\\  \small{of real disease-specific features.}\end{tabular}}  \\
\cline{2-4}
& \multicolumn{1}{c|}{\small{\textbf{P}} }& \multicolumn{1}{c|}{\small{50\%}}   & \small{\small{50\%}   }              &                                                                                                                                                                                      \\ \cline{2-4}
                                     & \multicolumn{1}{c|}{\small{\textbf{NP}}} & \multicolumn{1}{c|}{\small{50\%}}  & \small{50\%}                  &                                                                                                                                                                                                                                                                                                        \\ \hline
\end{tabular}
\label{tbl:test}
\end{table}

\section{Label-imbalance analysis setup}
For illustrating the validity of our hypothesis, we first describe our setup for generating label-imbalanced multi-source datasets followed by a description of the self-explainable model that is leveraged in this study.% in this section. 
\subsection{Datasets}
To consider a controlled setting, we take a balanced subset of classes with equal number of images from both the ChestX-ray14 and CheXpert datasets for Pneumonia detection. 

The NIH {\bf ChestX-ray14} dataset consists of 112,120 frontal-view X-Ray images from 14 classes \cite{chestxray14}. We split the data patient-wise into 80\% training, 10\% validation and 10\% test. For our controlled setting, we first select the Pneumonia class consisting of 1099 training images. We then sample 1099 images for the negative class randomly from the remaining 13 classes to ensure a balanced dataset and to remove the additional effect of class imbalance. Similarly, this results in 374 validation and 290 total test images.
We denote the images from this hospital system as H1.

{\bf CheXpert} is a large public dataset consisting of 224,316 chest radiographs of 65,240 patients consisting of 14 labels. \cite{chexpert}. Training and validation splits for this dataset are available. We separate the training split patient-wise into train and validation data and use the dataset's validation data for testing. To ensure two balanced datasets, we again sample 1099 images from the Pneumonia class and 1099 images from the negative class. We refer to the images from this hospital system as H2.

\subsubsection{Source induced label-imbalance}
\label{traindata}
In order to analyze the effect of label-imbalance across two different hospital systems (H1 and H2), we combine the two datasets such that they vary in their composition of Pneumonia and Non-Pneumonia images obtained from H1 and H2.
The substitution of the datasets is denoted as $x$H1 and $y$H2, where $x\in [0,100]$ denotes the percentage of Pneumonia training images taken from hospital system H1 and $y=100-x$ denotes the percentage of Pneumonia images taken from hospital system H2.
In total, we create 11 datasets and accordingly train 11 models, where $x$ takes a value in the range of 0-100 with an interval of 10. The Non-Pneumonia images are selected such as to always maintain the class balance, i.e $(100-x)$\% from H1 and $(100-y)$\% from H2. For example, in the case of $0$H1-$100$H2, the training data consists of 0\% Pneumonia images from H1, 100\% (1099) Pneumonia images from H2, 100\% (1099) Non-Pneumonia images from H1 and 0\% Non-Pneumonia images from H2. The validation data is selected following the same strategy of $x$H1-$y$H2. For testing, three configurations are used for all models to assess their behavior of recognizing source annotations (see Table \ref{tbl:test}).

\iffalse
\begin{table}[]
\centering
\begin{tabularx}{\columnwidth}{|X|X|}
\hline
\textbf{Name and Setup} & \textbf{Hypothesis}\\ \hline
\textbf{\small{Test-100H1:}}\small{100\% Pneumonia images are selected from H1, and 100\% Non-Pneumonia images are selected from H2.} & \small{If the model learns only hospital annotations instead of Pneumonia features, in the case of prevalence of Pneumonia from H1, the accuracy of this test dataset is expected to be 100\% and 0\% for the case of prevalence of Pneumonia from H2\%.}\\ \hline
\textbf{\small{Test-100H2:}} \small{100\% Non-Pneumonia images from H1 test data and 100\% Pneumonia images from H2.} & \small{The behavior of this setting is expected to be the opposite to that of Test-$100$H1.}\\ \hline
\textbf{\small{Test-50-50:}} \small{This setting takes 50\% of Pneumonia and 50\% of Non-Pneumonia images from both H1 and H2. } & \small{The accuracy of this test set for all the different models will indicate the learning of real disease-specific features by the models.}\\ \hline
\end{tabularx}
\caption{{...}}
\label{tbl:quant}
\end{table}
\fi

\subsection{Self-explainable method: PRP}
\label{sec:methods}
Considering the ability of black box models to learn spurious correlations \cite{chans, variable_generalization}, explainable AI is essential for automated medical image diagnosis.
Additionally, instead of explaining the black box models post-hoc, transparent self-explainable methods, which are capable of generating real-time explanations of the underlying decision process, can prove to be more faithful \cite{rudin2019stop}.
In this work, we leverage the recent method of PRP \cite{gautam2021looks} to obtain class-based prototypical explanation maps. 

PRP builds on ProtoPNet \cite{protopnet}, a self-explaining model, and consists of a class-specific prototype layer inserted between the convolutional output and the final fully connected layer. The convolutional output is denoted as $\mathbf{z} \in \mathcal{R}^{H\times W \times D}$, where $H$, $W$ and $D$ are the height, width and depth of $\mathbf{z}$. The prototype layer consists of a fixed number of prototypes per class, $\mathbf{P} = {\{\mathbf{p}_m\}_{m=1}^N}$ where $N$ are the total number of prototypes, each having a shape of $1\times 1 \times D$.  These are replaced by the closest training image patch features during training, thus representing each class by actual training image patches. $L_2$ similarities between the  prototypes $\mathbf{P}$ and patches of the convolutional output, $\tilde{\mathbf{z}}\in \text{patches}(\mathbf{z})$, of the input image are then computed to generate prototypical activation maps, $\mathbf{A} = {\{\mathbf{a}_m\}_{m=1}^N}$. This is followed by max pooling on the activation maps to generate the corresponding similarity scores $\mathbf{S} = {\{s_m\}_{m=1}^N}$. 
The network is trained in 3 steps: 1) training the whole network end to end, 2) projecting the prototypes to maintain explainability, i.e, replacing the prototypes by the convolutional output patch from the nearest training image of same class, 3) training the last layer. 

PRP, unlike ProtoPNet~\cite{protopnet}, does not perform visualization of relevant areas in the input via a model-agnostic bi-linear upsampling of the activation maps to the input size, but proposes a model-aware approach inspired by LRP \cite{lrp}. This leads to more faithful, higher resolution and spatially precise explanation maps by taking into account the network's structure and weights. Following PRP~\cite{gautam2021looks}, the relevance is distributed layer by layer to the input pixels, starting from similarity scores ($\mathbf{S}$).

\begin{figure}[!t]
    \centering
    \includegraphics[scale=0.25]{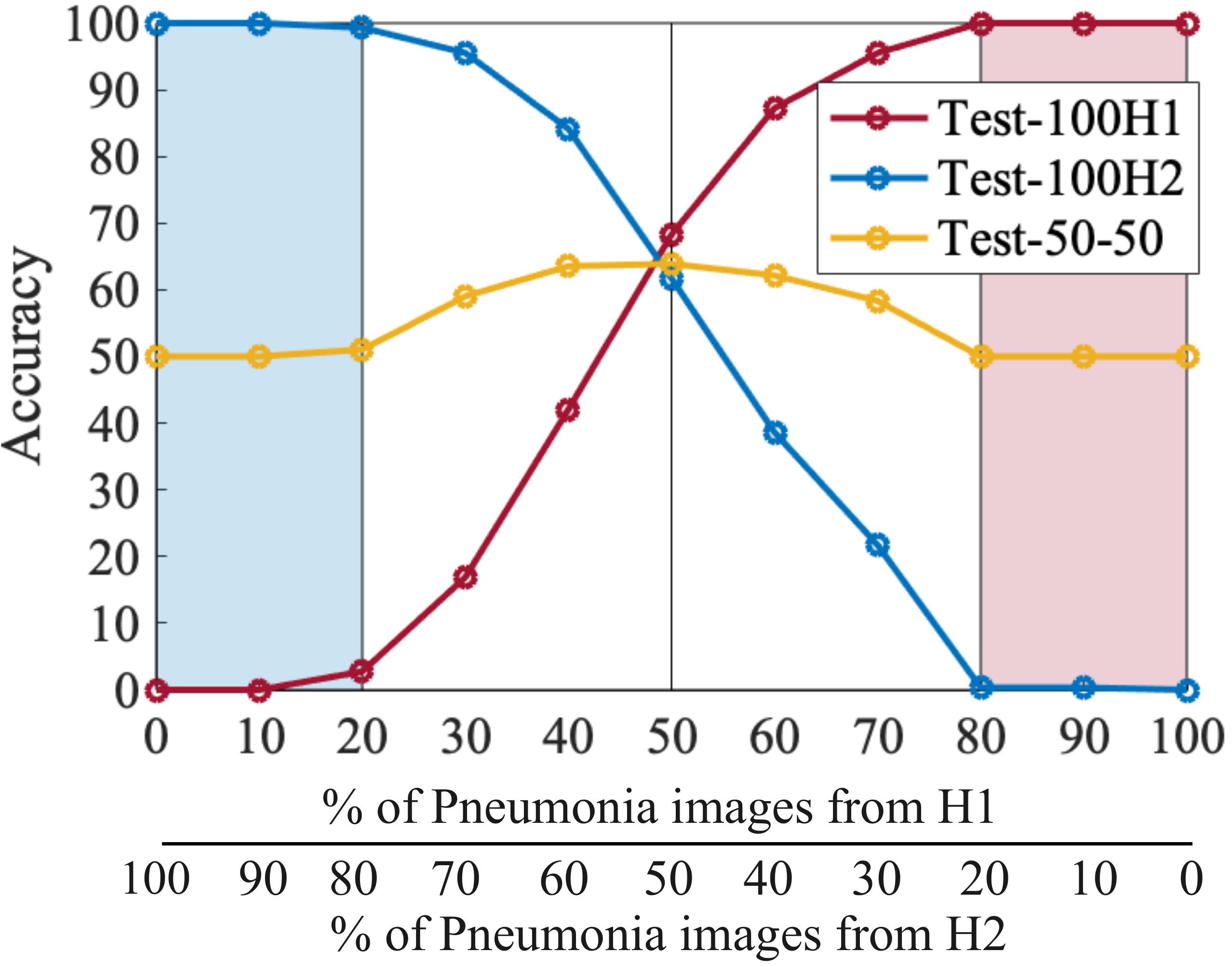}
    %\vspace{-2mm}
    \caption{{Accuracy for Test-100H1, Test-100H2 and Test-50-50 for different imbalances of Pneumonia images in training datasets based on the hospital systems. 11 models are trained by combining xH1 and yH2 images, where x (percentage of Pneumonia images from H1) is in the range of 0-100 (as shown on x-axis), with an interval of 10 and y=1-x is the percentage of Pneumonia images from H2.}}
    \label{fig:graph_1}
\end{figure}

\section{Experiments and Results}
We train 11 binary classification models with the architecture in \cite{gautam2021looks} for different dataset compositions as described in Section \ref{traindata}. Following \cite{protopnet}, the number of prototypes for each class is fixed to 10 and ResNet34 is used as the backbone. The network is trained for 50 epochs with a projection of prototypes after every 10 epochs, followed by training the last layer for 20 epochs.
The model selection is performed based on the best validation accuracy.

The test accuracies for Test-100H1, Test-100H2 and Test-50-50 for all models from 0H1-100H2 to 100H1-0H2 are shown in Fig. \ref{fig:graph_1}. The highest accuracy of 63.89\% for the Test-50-50 (yellow) is achieved by the model trained on 50H1-50H2.
Training even on a slight label-imbalanced dataset, 
% i.e, with a variation in the prevalence of Pneumonia from either one of the hospitals,
we can observe a significant decrease in the accuracy for Test-50-50.
%, thus indicating spurious learning. 
As we move from the center of the graph i.e, 50H1-50H2 towards the left, the percentage of Pneumonia images from H2 increases and H1 decreases. 
Consequently, the accuracy for Test-100H1 decreases and Test-100H2 increases, strengthening our hypothesis of cheating by the model by exploiting source information.
% spurious hospital related annotations for detecting Pneumonia.
In the blue shaded region on the left, the accuracy for Test-100H2 is almost 100\%, Test-100H1 is near 0\% and Test-50-50 is 50\%. This indicates that in the case of extreme label-imbalance, the model is only using the source related annotations for achieving better performance, thus acting as a hospital instead of a pneumonia detector.
The opposite observations can be made when moving from the middle to the right of the graph, as the percentage of Pneumonia images are increasing from H1.
The red region on the right mirrors the blue region, indicating the learning of only hospital based annotations by the models.

\begin{figure}[!t]
    \centering
    \includegraphics[scale=0.25]{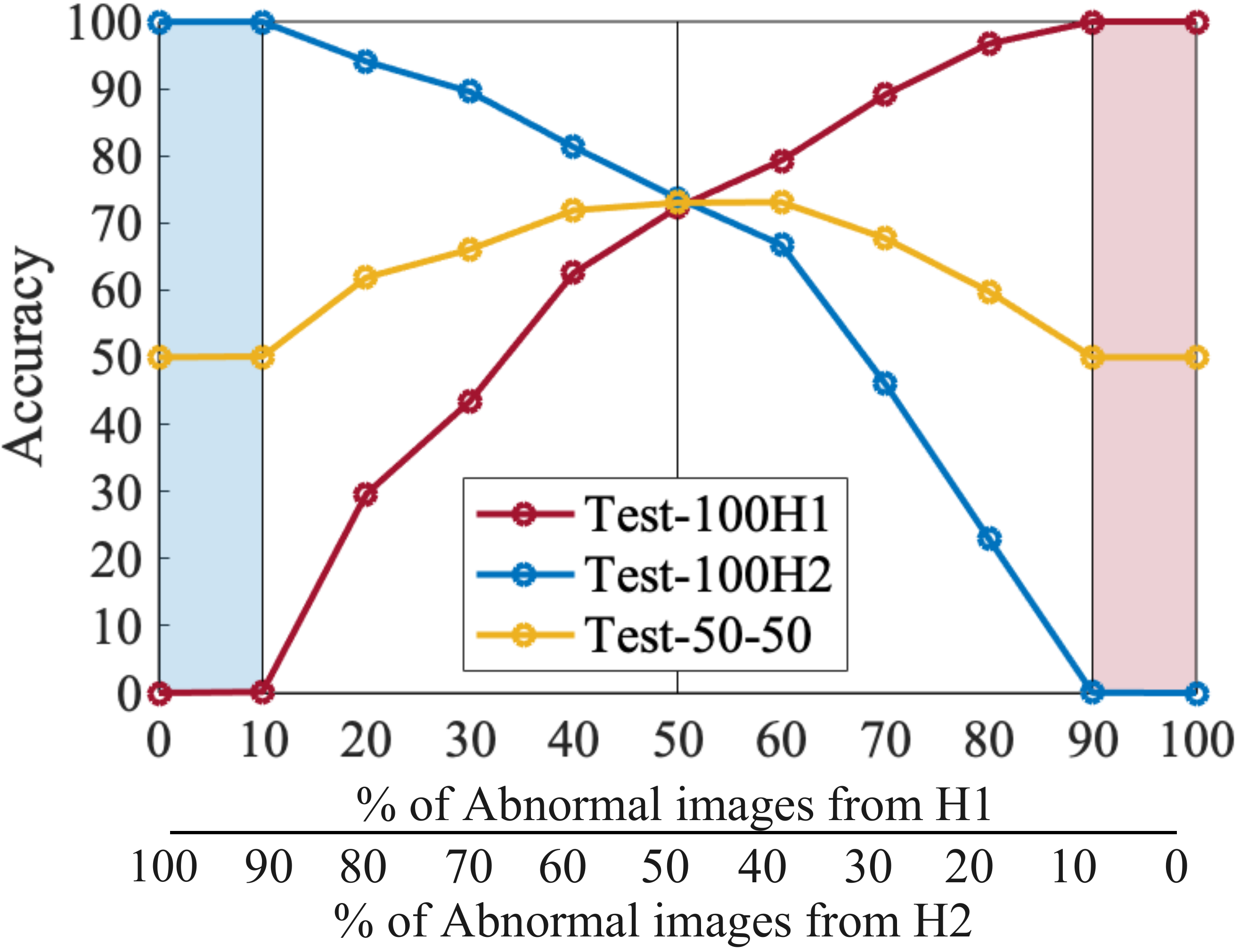}
    %\vspace{-2mm}
    \caption{{Accuracy for Test-100H1, Test-100H2 and Test-50-50 for gradual imbalance of Abnormal images based on the hospital systems. Learning of source-specific annotations can be observed while inducing even a slight label-imbalance.}}
    \label{fig:graph_2}
\end{figure}

Considering that Pneumonia detection is a difficult problem, we also repeat the same experiments for Abnormality detection. 
For this, we select data from the ``No Finding" category (absence of all pathologies) and data from the remaining 13 categories and consider them as the ``normal" and ``abnormal" class, respectively.
We again gradually induce label-imbalance by varying the percentage of abnormal data coming from H1 and H2 and train 11 models for this scenario. We follow the same setting for the test configurations (see Table \ref{tbl:test}) where the percentages now correspond to normal and abnormal class data.
%The test sets consist of the same setting as in the Pneumonia detection problem. 
The results are shown in Fig. \ref{fig:graph_2}, where the accuracy of the Test-50-50 goes up to 73.01\% when using the label-balanced training data 50H1-50H2. As the imbalance increases i.e, moving to the left or right of the graph from the middle, the models start behaving as a hospital detector. Although, the effect is less severe than for the more difficult problem of Pneumonia detection, hospital detection is still observed in all cases except for the label balanced dataset of 50H1-50H2. This further stresses the importance of using label balance data for multi-source data analysis. 
\begin{figure}[!t]
    \centering
    \includegraphics[scale=0.45]{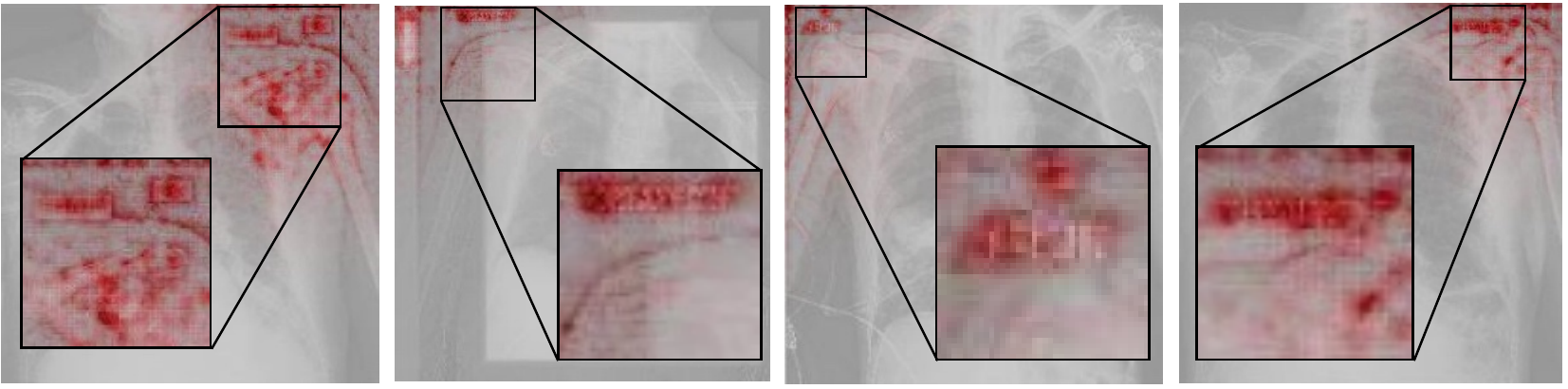}
    \vspace{-2mm}
    \caption{{Visualization of the PRP maps for the top four activated images (having the highest similarity score) by the Non-Pneumonia class prototype for 90H1-10H2 shown in Fig. \ref{fig:tag_prototype}(blue). Note, this prototype is able to capture several different kinds of annotations from hospital system H2.}
    }
    \label{fig:top3}
\end{figure}

\begin{figure}[!tpb]
    \centering
    \includegraphics[scale=0.45]{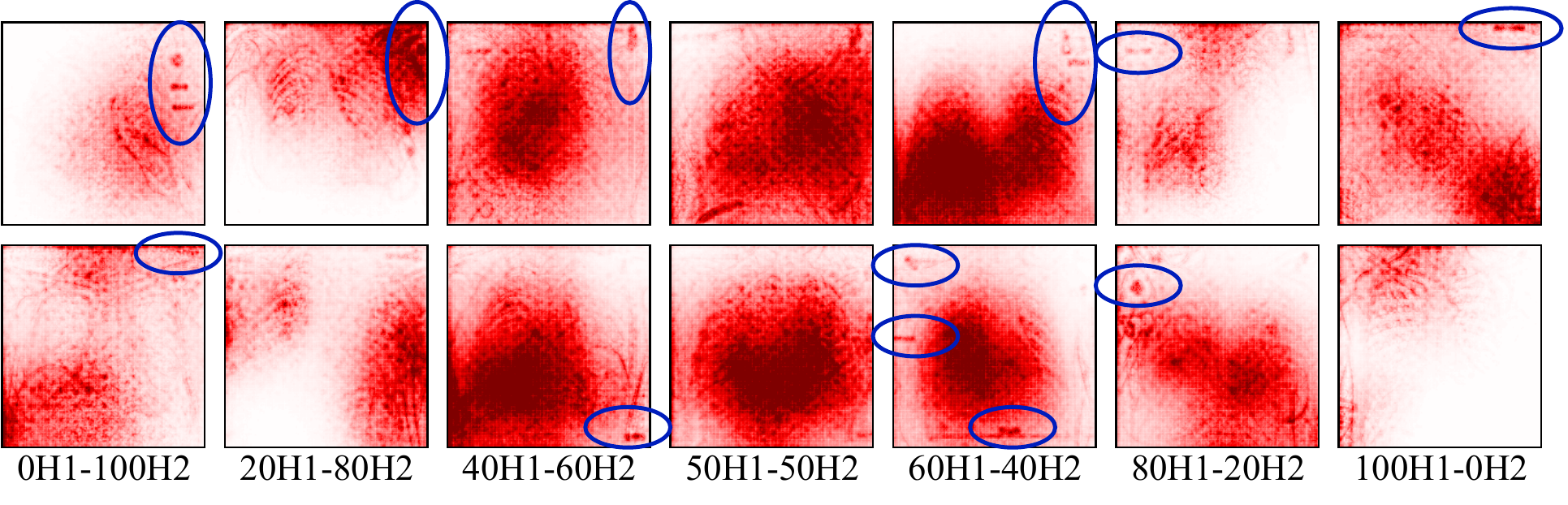}
    \vspace{-7mm}
    \caption{{Global PRP maps for Non-Pneumonia class in row 1 and Pneumonia class in row 2 for different models with learnt annotations marked in blue. As the imbalance in the hospital decreases, the models focus more on the center of the image and less on the annotations.}}
    \label{fig:global}
\end{figure}

\begin{figure}[!t]
    \centering
    \includegraphics[scale=0.65]{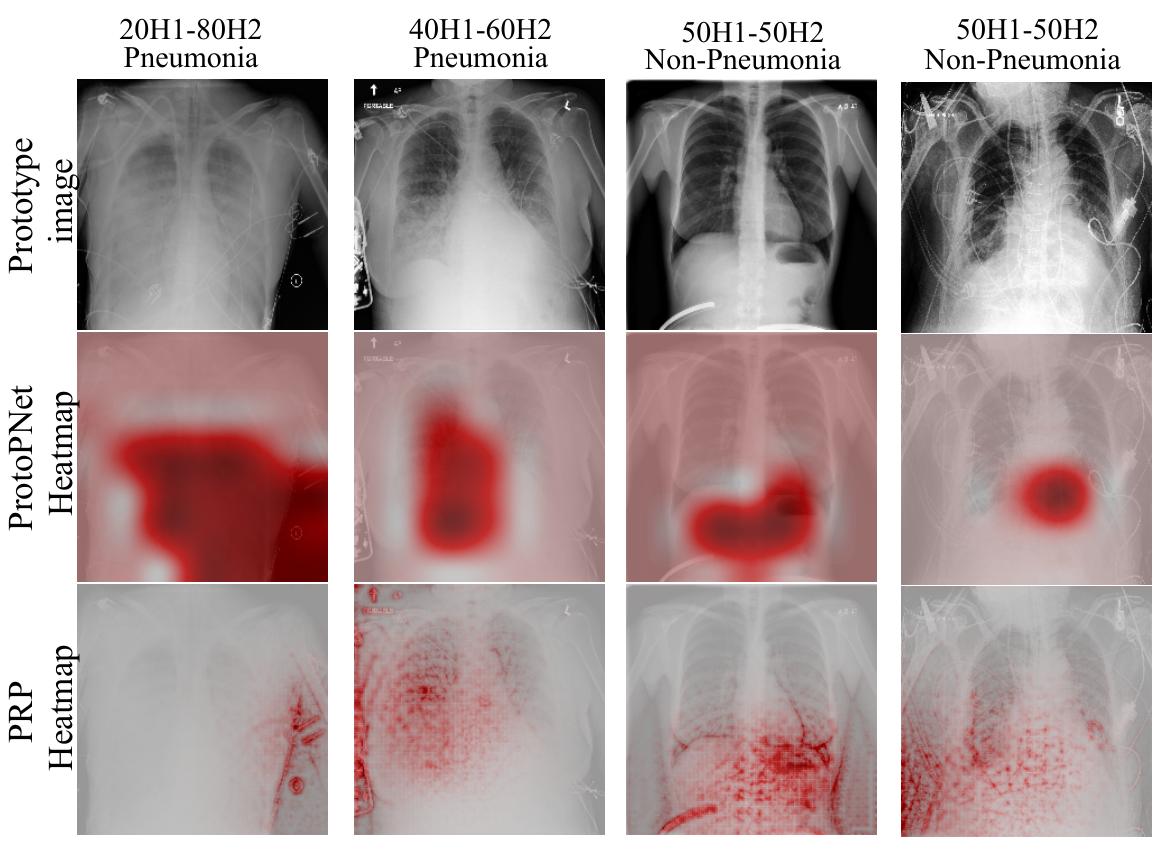}
    \vspace{-3mm}
    \caption{{Other artifacts captured by the prototypes. The rows represent the original image where the prototype comes from, ProtoPNet explanation heatmap and PRP map, respectively. Model names and prototype class are mentioned on the top.}}
    \label{fig:other_arttifacts}
\end{figure}

To demonstrate the significance of self-explaining models for detecting spurious artifact learning, for a prototype from the Non-Pneumonia class of 90H1-10H2 model (Fig. \ref{fig:tag_prototype}, blue), we visualize the top 4 activated training images by this prototype in Fig. \ref{fig:top3}. The PRP maps, overlayed on the images, clearly indicate the activation of various kinds of source annotations, which are captured by the prototype, as shown in the zoomed regions.

To visualize the aggregate information learned for a class, for each model, the PRP maps for all unique prototypes learned for both, Pneumonia and Non-Pneumonia class are 
% aggregated as union
superimposed 
and shown in Fig. \ref{fig:global}, thus representing the class-specific global prototypes. The annotations learned by the models are circled in blue. Interestingly, as we move from 50H1-50H2 to even a slight imbalance of 60-40, the models start capturing the textual annotations. From left to right in Fig. \ref{fig:global}, it can be observed that the models start focusing more in the center of the image, trying to capture the real disease-specific features when the imbalance decreases, especially in the case of 50H1-50H2, where the global PRP maps strongly highlight the center of the images for both the classes.

In Fig. \ref{fig:other_arttifacts}, we visualize the PRP maps for various prototypes capturing, among others, spurious information in the images. From the PRP maps we can observe that the model is capturing artifactual medical instruments, such as chest tubes, drips, and glucose bottles. It is interesting to note that in the label-balanced case, even if the reliance on source-specific artifacts is reduced, the model can still capture disease-specific artifacts in addition to the pathology features to achieve better performance (Fig. \ref{fig:other_arttifacts} column 3 and 4). An example of this can be the prevalence of chest tubes in Pneumothorax class \cite{variable_generalization}, which are being captured by Non-Pneumonia class prototype in column 4 of Fig. \ref{fig:other_arttifacts}, thus inaccurately diagnosing the absence of chest tubes as Pneumonia. These observations further stress on the importance of self-explainable models even for the label-balanced datasets. Additionally, to visualize the strengths of PRP over ProtoPNet, we also show the corresponding ProtoPNet heatmaps in Fig. \ref{fig:other_arttifacts} in the middle row. As can be seen the PRP maps are more precise 
as opposed to the inconclusive and coarse ProtoPNet explanations. 

For quantifying the faithfulness of PRP maps over ProtoPNet heatmaps, we calculate the Average Drop (A.D.) and Average Increase (A.I.) with respect to similarity scores corresponding to the predicted class prototypes. Following \cite{Lee_2021_CVPR}, we mask out the 50\% least activated pixels in the heatmaps replacing them with random uniformly sampled values.
%use the most activated 50\% of the pixels of the heatmaps as masks. For generating masked images, we replace the background pixels with random uniform values. 
A.D is then expressed as $\sum_{m=1}^n \sum_{i=1}^K \frac{max(0, s_m(i) - o_m(i))}{s_m(i)}\times \frac{100}{n\times K} \;$,
% \begin{equation}
%     \sum_{m=1}^n \sum_{i=1}^K \frac{max(0, s_m(i) - o_m(i))}{s_m(i)}\times \frac{100}{n\times K} \; ,
% \end{equation}
where $s_m(i)$ is the similarity score for prototype $m$ and image $i$, $o_m(i)$ is the output similarity score for the masked image and $n$, $K$ are the number of prototypes for the predicted class and total number of images, respectively. A.I is expressed as $\sum_{m=1}^n \sum_{i=1}^K \mymathbb{1}[s_m(i) < o_m(i)] \times \frac{100}{n\times K} \; ,$
% \begin{equation}
%     \sum_{m=1}^n \sum_{i=1}^K Sign( s_m(i) < o_m(i)) \times \frac{100}{n\times K} \; ,
% \end{equation}
where $\mymathbb{1}[\cdot]$ is the Iverson bracket indicator function that returns 1 when the condition is true. Table \ref{tbl:quant} shows the values for A.I. and A.D. for both ProtoPNet and PRP maps, averaged over all images in the corresponding test sets and prototypes for models 0H1-100H2, 50H1-50H2 and 100H1-0H2 with test sets Test-100H2, Test-50-50 and Test-100H1, respectively. The results demonstrate that PRP performs consistently better or comparable to ProtoPNet for generating more faithful explanations.

\renewcommand{\tabcolsep}{8pt}
\begin{table}[]
\def\arraystretch{0.9}
\centering
\caption{{AI and AD for similarity scores of predicted class prototypes on corresponding test sets for the different models i.e, Test-100H2, Test-50 and Test-100H2 for 0H1-100H2, 50H1-50H2 and 100H1-0H2, respectively. Lower A.D. and Higher A.I suggests better performance.}}
\begin{tabular}{|c|c|c|c|c|}
\hline
                   & \multicolumn{2}{c|}{\textbf{\small{A.D.}}} & \multicolumn{2}{c|}{\textbf{\small{A.I.}}} \\ \hline
                   & \textit{\small{ProtoPNet}}  & \textit{\small{PRP}} & \textit{\small{ProtoPNet}}  & \textit{\small{PRP}} \\ \hline
% \textbf{\small{\rotatebox[origin=c]{90}{\parbox[c]{1cm}{\centering 0H1-100H2}}}}
\textbf{\small{0H1-100H2}}&    \small{13.07}         &    \small{9.70}        &    \small{38.44}         &    \small{43.58}         \\ \hline

\textbf{\small{50H1-50H2}} &  \small{52.58}             &      \small{48.37}     &           \small{20.06}      &   \small{22.08}       \\ \hline
% \textbf{100H1-0H2} &    12.46                 &    12.48          &        48.36             &        50.83  \\ \hline
\textbf{\small{100H1-0H2}} &    \small{12.47}          &    \small{12.92}        &        \small{45.65}          &        \small{48.10}  \\ \hline
\end{tabular}
\label{tbl:quant}
\end{table}

\section{Conclusion}
Multi-source applicability of black box deep learning models remains questionable. In this work, we demonstrate that the models are prone to learning spurious correlations in terms of textual annotations for Chest X-Ray image analysis in the presence of source induced label-imbalances. Even with a slight imbalance, the models are inclined to cheat and act as a hospital detector instead of the disease detector. Consequently, we recommend to ensure label-balancing while using multi-source datasets for efficient clinical deployment. Further, using a self-explainable method of PRP, we highlight the importance of using more transparent self-explainable models for real-time detection of spurious learning. \\
\textbf{
Compliance with Ethical Standards:} No ethical approval was required because of the retrospective use of open source datasets.
\bibliographystyle{IEEEbib}
\bibliography{ref}

\end{document}